\documentclass[doublecol]{epl2}

\usepackage{graphicx}
\usepackage{latexsym}
\usepackage{amsmath}

\newcommand{\Rgyr}{\mbox{$R_\text{g}$}}

\newcommand{\bg}{\mbox{$a$}}

\newcommand{\cstar}{\mbox{$c^*$}}
\newcommand{\Ustar}{\mbox{$U^*$}}

\title{Why polymer chains in a melt are not random walks}
\shorttitle{Why polymer chains in a melt are not random walks}

\author{J.~P. Wittmer\inst{1}\thanks{E-mail: jwittmer@ics.u-strasbg.fr} \and
        P.~Beckrich\inst{1} \and
        A.~Johner\inst{1} \and
        A.~N.~Semenov\inst{1} \and
        S.~P.~Obukhov\inst{1,2} \and
        H.~Meyer\inst{1} \and
        J.~Baschnagel\inst{1} 
}
\shortauthor{J.P. Wittmer \etal}

\institute{
 \inst{1} 
 Institut Charles Sadron, 6 Rue Boussingault, 67083 Strasbourg Cedex, France\\
 \inst{2} 
 Department of Physics, University of Florida, Gainesville FL 32611, USA
}


\abstract{
A cornerstone of modern polymer physics is the `Flory ideality hypothesis'
which states that a chain in a polymer melt adopts `ideal' 
random-walk-like conformations.
Here we revisit theoretically and numerically this pivotal assumption
and demonstrate that there are noticeable deviations from ideality. The 
deviations come from the interplay of chain connectivity and the
incompressibility of the melt, leading to an effective repulsion between 
chain segments of all sizes $s$. 
The amplitude of this repulsion increases with decreasing $s$ where 
chain segments become more and more swollen.
We illustrate this swelling by an analysis 
of the form factor $F(q)$, i.e.\ the scattered intensity at wavevector $q$ 
resulting from intramolecular interferences of a chain. A `Kratky plot' of 
$q^2F(q)$ {\em vs.} $q$ does not exhibit the plateau for intermediate 
wavevectors characteristic of ideal chains. One rather finds a 
conspicuous depression of the plateau, $\delta(F^{-1}(q)) = |q|^3/32\rho$, 
which increases with $q$ and only depends on the monomer density $\rho$. 
}

\begin{document}
\maketitle

Polymer melts are dense disordered systems consisting of macromolecular chains.
Theories that predict properties of chains in a melt or concentrated solutions
generally start from the `Flory ideality hypothesis' \cite{FloryBook}. The 
hypothesis states that polymer conformations correspond to those of `ideal' 
random walks on length scales much larger than the monomer diameter
\cite{FloryBook,DegennesBook,DoiEdwardsBook}.
The commonly accepted justification is that {\em intra}chain 
and {\em inter}chain excluded volume forces compensate
each other in dense systems \cite{DegennesBook,DoiEdwardsBook}. 
This compensation has several important consequences.  For instance, the radius of 
gyration $R(s)$ of chain segments of curvilinear length $s \le N$ scales as 
$R^2(s) = \bg^2 s$, where $N$ denotes the number of monomers per chain and $a$ 
the statistical segment length of the chain.  This result holds provided 
$s$ is sufficiently large for all local correlations to be neglected. For $s=N$, it 
implies that the radius of gyration of the total chain obeys
$\Rgyr = R(s=N) = \bg \sqrt{N}$ \cite{DoiEdwardsBook}. 
A further consequence of chain ideality is that the intrachain scattering function, 
the `form factor' $F(q)$, is given by the Debye formula, 
$F^{(0)}(q) = 2 N \left(\exp(-(q\Rgyr)^2) -1 + (q\Rgyr)^2 \right)/(q\Rgyr)^4$
 \cite{DoiEdwardsBook}. For intermediate wavevectors, $1/\Rgyr \ll q \ll 1/a$,
$F^{(0)}(q)$ reduces to the power law $F^{(0)}(q) \approx 2/(q \bg)^2$. Due to this
power-law behavior, we refer to the latter $q$-regime as `scale-free regime' in the following. 

\begin{figure}[t]
\onefigure[width=0.49\textwidth]{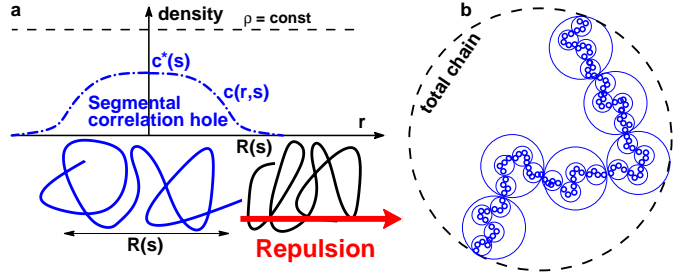}
\caption{ 
Role of incompressibility and chain connectivity in dense polymer solutions.
{\bf (a)} 
Sketch of the segmental correlation hole of a marked chain segment of 
curvilinear length $s$. Density fluctuations of chain segments are correlated, 
since the total density (dashed line) cannot fluctuate in dense polymer solutions.
{\bf (b)}
Self-similar pattern of nested segmental correlation holes aligned along the 
backbone of a reference chain. The large dashed circle represents the classical 
correlation hole of the total chain ($s=N$) \cite{DegennesBook}. 
Here we argue that incompressibility and chain connectivity lead to a repulsion 
of the segmental correlation holes, which increases with decreasing $s$. 
The ensuing swelling of chain segments manifests itself 
by deviations of the form factor from Debye's formula.
}
\label{fig_sketch}
\end{figure}

\begin{figure}[t]
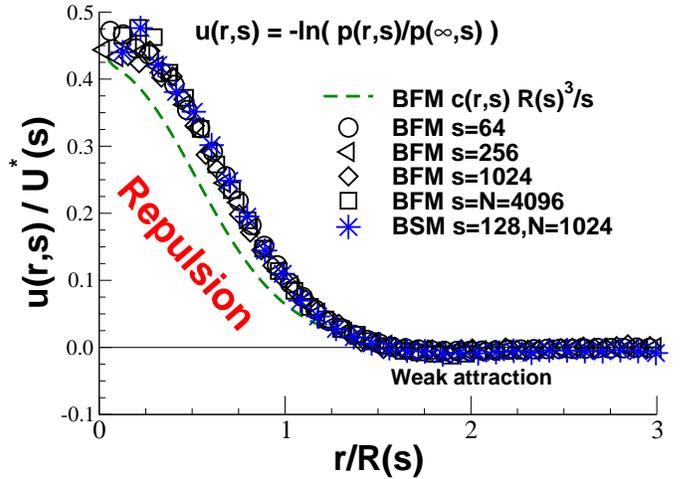

\onefigure[width=0.49\textwidth]{fig_2}
\caption{
%
Pair-distribution function $p(r,s)$ of the centers of mass of chain segments of length $s$.
Data for different chain and segment lengths are successfully scaled by 
plotting $u(r,s) / \Ustar(s)$ as a function of 
$r/R(s)$ with $\Ustar(s) \equiv s/\rho R(s)^3$.
The figure compares different $s$ for BFM chains of length $N=4096$ and one BSM 
data set for $N=1024$ (stars). The function is positive 
for small $r/R(s)$ where segments repel each other. 
It is weakly attractive for larger distances. 
%
%
For comparison we also indicate the self-density $c(r,s) R(s)^3/s$ for one BFM
example with $s=256$ (dashed line) which is similar --- albeit not identical ---
to $u(r,s)/\Ustar(s)$. 
}
\label{fig_Uspot} 
\end{figure}

\begin{figure}[t]
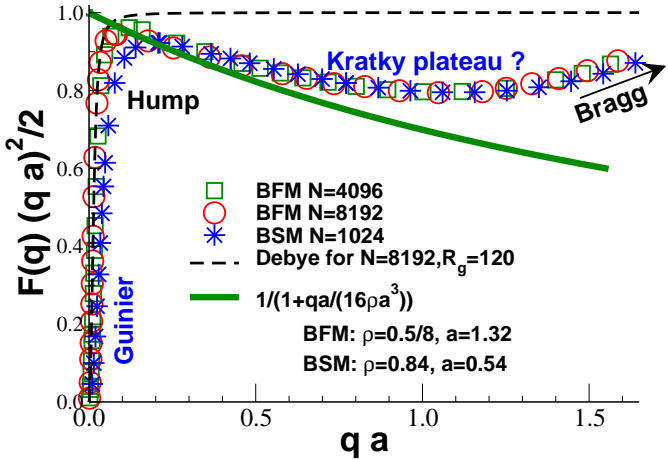

\onefigure[width=0.49\textwidth]{fig_3}
\caption{Kratky representation of the form factor $F(q) (q \bg)^2/2$ 
{\em vs.} rescaled wavevector $q \bg$.
The axes are rescaled by the statistical segment length \bg \ for the BFM and BSM 
(as indicated in the figure) to make both data sets readily comparable. The values 
of \bg \ for asymptotically long chains were obtained by a fit to 
eq.~(\ref{eq_RsUs}). Note that $\rho \bg^3 \approx 0.13$ is similar for both models
which explains the collapse for small $q$. 
(The agreement for large $q \bg$ is fortuitous.)
The Debye formula (dashed line), computed here for $N=8192$ using the measured 
radius of gyration ($\Rgyr\approx 120$), overestimates the dip of the form factor 
at $q \bg \approx 1$ by about $20\%$.
%
%
The bold line indicates the deviation from the Kratky plateau, 
eq.~(\ref{eq_keyclaim}), with the choice $F^{(0)}(q) \ (q \bg)^2/2 = 1$, 
valid for ideal chains in the scale-free $q$-regime.
The increase of $F(q)$ for the largest wavevectors shown is due to non-universal 
correlations on the monomer scale. Such microstructure effects are not captured by 
the theory.
}
\label{fig_Fkrat}
\end{figure}

Neutron scattering experiments have been extensively used to test Flory's 
hypothesis \cite{BenoitBook}. This technique allows one to extract $F(q)$ from the 
total scattered intensity of a mixture of deuterated and hydrogenated polymers. 
One usually plots $q^2 F({q})$ versus $q$ (`Kratky plot') to reveal the existence 
of a `Kratky plateau' in the scale-free regime. In applications, 
however, the Kratky plateau appears to be elusive \cite{BenoitBook}.
Possible causes for deviations are effects of chain stiffness and finite chain 
thickness. In special cases, the scattering signals from chain stiffness and 
thickness may compensate one another, leading fortuitously to an extended Kratky 
plateau \cite{MRRDCP1987,Higgins}. Apparently, `Kratky plots have to be 
interpreted with care' \cite{BenoitBook}. 
It is generally believed that if all possible obscuring factors are avoided and 
`thin' flexible polymers are examined, no deviations from the Kratky plateau 
should occur. Here we show that this is not true because long-range correlations 
along the chain backbone are induced by repulsive interactions of chain segments 
in dense polymer systems \cite{SJ03,WMBJOMMS04}. 
This effect is related to the well-known correlation hole \cite{DegennesBook}. For 
chain segments of length $1 \ll s < N$, the correlation hole leads to non-Gaussian
deviations of the form factor. Our key claim is that for asymptotically long chains 
($N\rightarrow \infty$) the deviations are given in the scale-free regime by
\begin{equation}
\delta\left(\frac{1}{F(q)}\right) \equiv 
\frac{1}{F(q)} - \frac{1}{F^{(0)}(q)} = \frac{1}{32} \frac{|q|^3}{\rho}
\label{eq_keyclaim}
\end{equation}
with $\rho$ being the monomer number density.
In the following, we first outline the theory (sketched in fig.~\ref{fig_sketch}) 
and provide a scaling argument for eq.~(\ref{eq_keyclaim}). Computer simulations 
of two well-studied coarse-grained polymer models, both described in detail in 
Ref.~\cite{BWM04}, are then utilized to carry out a critical test of our 
predictions (figs.~\ref{fig_Uspot}, \ref{fig_Fkrat} and \ref{fig_dFqRg}). 

The general physical idea of our theory and the long-range intrachain correlations 
it predicts is related to the `segmental correlation hole' of a chain segment of 
length $s$ in $d$ dimensions (fig.~\ref{fig_sketch}). 
Polymer melts are essentially incompressible (on length scales large compared
to the monomer diameter), and the density $\rho$ of {\em all} monomers does not 
fluctuate. On the other hand, composition fluctuations of marked chains or 
subchains may occur, subject to the constraint of constant density.
Figure~\ref{fig_sketch} shows a marked chain segment of curvilinear length $s$ 
within a much larger chain of length $N \gg s$. The monomer density distribution 
$c(r,s)$ of the segment (dash-dotted line) becomes 
$\cstar = c(r\approx 0,s) \approx s/R(s)^d$ close to the center of mass ($r=0$) 
and decays rapidly at distances of order $R(s)$. 
Since composition fluctuations are coupled by the density constraint, 
a second chain segment (of the same or another chain) feels an {\em entropic} 
penalty $\Ustar(s) \approx \cstar(s)/\rho$
when both correlation holes approach each other \cite{SJ03}. 
(The scaling of $\Ustar(s)$ is demonstrated below, fig.~\ref{fig_Uspot}.) 
This repulsion when considered between two half-segments swells the segment.
In $d=3$, the effect is weak, $\Ustar(s) \equiv s/\rho R(s)^3 \sim 1/\sqrt{s}$, 
and a standard perturbation calculation can be performed 
\cite{DoiEdwardsBook,Beckrich}.
This calculation considers quantities which are defined such that they vanish if 
the perturbation potential $\Ustar$ is switched off, and are then shown to depend, 
in leading order, linearly on $\Ustar$ (which may be directly checked numerically).
For instance, one finds that \cite{WMBJOMMS04}
\begin{equation}
1- \frac{R^2(s)}{\bg^2 s} \approx \Ustar(s)
\label{eq_RsUs}
\end{equation}
describes the swelling of chain segments in real space.
Note that $\bg$ is the renormalized statistical segment
characterizing infinite chains in the melt. 
In reciprocal space, this swelling appears in the form factor.
As the segment length is related to the wavevector $q$ and the ideal
form factor $F^{(0)}(q)$ by $s(q) \sim 1/|q|^2 \sim F^{(0)}$,
we find for the deviations of the measured and ideal form factors
\begin{equation}
\delta \left(\frac{1}{F(q)}\right) =
\frac{1}{F^{(0)}} \times \left(\frac{F^{(0)}}{F} - 1 \right)
\approx \frac{1}{s} \times \Ustar(s) \approx \frac{q^d}{\rho}
\label{eq_dFscal_s}
\end{equation}
which agrees with eq.~(\ref{eq_keyclaim}). The important feature of this 
$|q|^3$-correction is that it depends neither on the strength of the excluded 
volume interaction nor on the statistical segment length. Hence, it must be 
generally valid, even for semidilute solutions. We checked explicitly that the 
result from the renormalization group theory for semidilute solutions 
\cite{SchaferBook,SMK00} takes the same form as eq.~(\ref{eq_keyclaim}) with the 
amplitude $0.03124\ldots$ that is within $0.03\%$ of our $1/32$ \cite{Beckrich}.

The scaling of the effective interaction between two chain segments
can be tested numerically by computing the correlation hole function
$u(r,s) \equiv - \ln(p(r,s)/p(\infty,s))$ where $p(r,s)$ denotes the 
pair-distribution function of the centers of mass of all segments of length $s$.
Since the correlation hole is shallow for large $s$, expansion leads to 
$u(r,s) \approx 1 - p(r,s)/p(\infty,s) \approx c(r,s)/\rho$ \cite{SJ03}.
Hence the interaction strength at $r/R(s) \ll 1$ should be given by
$\Ustar(s) \sim u(0,s) \sim \cstar(s)/\rho$.
Figure~\ref{fig_Uspot} shows that a scaling in terms of the measured radius of 
gyration $R(s)$ and $\Ustar(s)$ is indeed possible. This finding confirms that 
segments repel each other for small $r/R(s)$ with an entropic penalty 
$\Ustar(s) \sim 1/\sqrt{s}$ for large $s \gg 10$.\footnote{Strictly speaking, 
$u(r,s)$ coincides with the effective interaction 
potential $\Ustar(r,s)$ of two {\em independent} segments of length $s$ only if $s=N$. 
More generally, $u(r,s)$ reflects also the interactions of their neighbors 
along the chains. However, this does not affect the scaling
on short distances, $r \le R(s)$, which matters here.}

%
%
Here, as in the subsequent figures, the body of our data comes from the bond 
fluctuation model (BFM) --- a lattice model which we study by Monte Carlo (MC) 
simulations. Using a mixture of local, slithering snake, and double-bridging MC 
moves \cite{BWM04} we were able to equilibrate dense systems ($\rho=0.5/8$) with 
chain lengths up to $N=8192$ contained in periodic simulation boxes of linear size 
$L=256$. Additionally, molecular dynamics simulations of a bead-spring model (BSM) 
\cite{BWM04} were performed to dispel concerns that our results are influenced by 
the underlying lattice structure of the BFM.  For clarity, we show only one BSM 
data set ($N=1024$, $\rho \approx 0.84$, $L \approx 62$).
All length scales are given in units of the lattice constant or the bead diameter 
for the BFM or BSM, respectively. 
Compared to standard experimental polymers both models have
in common that the chains are very flexible and the units (beads)
are point-like.
They thus represent an ideal tool for testing the theory.

Figure~\ref{fig_Fkrat} presents a Kratky plot of $F(q)$ obtained for the 
longest chains currently available for both models. It reveals clear 
deviations from the ideal form factor (dashed line).
Instead of reaching a plateau, the simulated $F(q)$ continuously decreases with 
increasing $q$ in the scale-free regime. This difference is expected from our 
theory. The result for infinitely long chains ---
eq.~(\ref{eq_keyclaim}) with $1/F^{(0)}(q) = (q\bg)^2/2$ (bold line) ---
enables us to understand the initial depression below the Kratky plateau. 
The prediction for infinite $N$ cannot capture the decrease of the 
form factor for $qa \le 0.1$ leading to the Guinier regime \cite{BenoitBook} 
where $F(q)$ is determined by the finite size of the simulated chains. A clearer 
evidence for the theory should thus be obtained by a different comparison between 
theory and simulation, which accounts for finite-$N$ effects. 
\begin{figure}[t]
\onefigure[width=0.48\textwidth]{fig_4}
\caption{Scaling of the form factor deviations 
$\delta\left( N/F(q) \right) /\Ustar(N)$ plotted {\em versus} $Q= q \Rgyr(N)$
as suggested by eq.~(\ref{eq_FNscaling}). Here $\Ustar(N)=N/\rho R_\mathrm{g}^3(N)$ 
and $\delta\left(1/F(q) \right) = 1/F(q) - 1/F^{(0)}(q)$, where $F(q)$ is the 
measured form factor and $F^{(0)}$ denotes the Debye function. For each data set 
the scaling of the axes and the calculation of the Debye function are done with the
measured radius of gyration $\Rgyr(N)$. Without free parameter perfect data collapse
is obtained for all systems. Deviations from data collapse are expected for large 
$Q$ due to finite persistence length and microstructure effects (`Bragg' regime).
In the Guinier regime ($Q \ll Q_0$), $\delta\left(1/F(q) \right)$ increases rapidly
as $Q^4$ (thin line). Most importantly, the {\em non-analytic} $|Q|^3$-power law 
slope (bold line) can be seen over an order of magnitude in $Q$
--- a striking confirmation of the theory. 
Obviously, the larger the chains, the better the agreement
with eq.~(\ref{eq_keyclaim}) which is valid for asymptotically long chains only.
}
\label{fig_dFqRg}
\end{figure}

This is achieved in fig.~\ref{fig_dFqRg} which focuses on deviations 
$\delta(1/F(q))=1/F-1/F^{(0)}$ from the Debye formula $F^{(0)}$.
If the axes are rescaled according to 
\begin{equation}
m(Q) \equiv \delta (N/F(q)) / \Ustar(N) 
\label{eq_FNscaling}
\end{equation} 
with $Q \equiv q \Rgyr(N)$,
a perfect collapse of the simulation data for all $N$ and both polymer models is 
obtained by using only one length scale, the (independently) measured $\Rgyr(N)$.
Eq.~(\ref{eq_FNscaling}) is suggested by the fact that $F(q)$ is proportional to 
the number of scatterers, $N$, and is a function of $q\Rgyr(N)$ only 
\cite{DoiEdwardsBook}. The master curve $m(Q)$ thus generalizes 
eq.~(\ref{eq_keyclaim}) to finite $N$.
In the Guinier regime ($Q \ll Q_0 \approx 6.3$) we find $m(Q) \sim Q^4$, 
as expected from a standard {\em analytic} expansion in $Q^2$.
(The first two terms in $Q^0$ and $Q^2$ must vanish by construction.)
The scale-free regime is observed for $Q > Q_0$. Since $\delta (1/F) \sim N^0$ 
in this regime, it follows from eq.~(\ref{eq_FNscaling}) that $m(Q) \sim |Q|^3$,
which provides an additional scaling argument for eq.~(\ref{eq_keyclaim}).
Fig.~\ref{fig_dFqRg} demonstrates for the largest chains available over more than 
an order of magnitude in $Q$ a good quantitative agreement (bold line).
We therefore suggest that an analysis of experimental data should employ this 
scaling plot as a diagnostic for the effects discussed here.

In summary, we have identified a general mechanism that gives rise to deviations 
from ideal chain behavior in polymer melts, even if the chains are very flexible 
and thin. This mechanism rests upon the interplay of chain connectivity and the 
incompressibility of the system, which generates an effective repulsion between 
chain segments of curvilinear length $s$ (fig.~\ref{fig_sketch}). A polymer in 
dense solutions may not be viewed as {\em one} soft sphere (or ellipsoid) 
\cite{Likos,Guenza}, but as a hierarchy of nested segmental correlation holes of 
all sizes aligned and correlated along the chain backbone.  
The repulsion scales like $\Ustar(s) \sim 1/\sqrt{s}$ for $s \gg 1$ 
(fig.~\ref{fig_Uspot}); it is strong for small $s$, but becomes weak for 
$s\rightarrow N$ in the large-$N$ limit.\footnote{Consequently, 
a contribution $\Ustar(N) \sim 1/\sqrt{N}$ to the free energy 
of a polymer chain in the melt exists scaling {\em non-linearly} with molecular weight.
This may be demonstrated computationally from 
the size-distribution in systems of linear self-assembled equilibrium polymers. 
}
The overall size of a long chain thus remains almost `ideal', whereas subchains 
are swollen (eqs.~(\ref{eq_keyclaim},\ref{eq_RsUs})). This swelling manifests 
itself in the form factor. In the Kratky representation, $F(q)$ displays a hump 
followed by a decrease in the $q$-regime where the Debye function would exhibit a
plateau (fig.~\ref{fig_Fkrat}). 
It should be stressed that eq.~(\ref{eq_keyclaim}) does not predict an 
{\em analytic} expansion in terms of $q^2$ as one might naively anticipate. 
The intriguing $|q|^3$-asymptote for infinite chains (fig.~\ref{fig_dFqRg}) 
formally arises from dilation invariance of the calculated diagrams \cite{Beckrich}. 
Established theoretical methods \cite{Curro,Fuchs} 
{\em assuming} implicitly analytical 
properties of scattering functions must therefore overlook these non-analytical terms.

These deviations from ideality should be measurable by neutron scattering 
experiments of {\em flexible} polymers. Unfortunately, finite persistence length 
effects may mask the predicted behavior if the chains are not sufficiently long.
An experimental verification 
--- e.g. following the lines of a promising recent study \cite{Higgins} ---
would be of great fundamental interest; it could also delineate the conditions where 
the predicted corrections to ideality can be observed in real polymer systems and must 
be considered in understanding their structure, phase behavior, and dynamics. 
Such an impact on structure and dynamics is expected theoretically, for the bulk 
\cite{WMBJOMMS04,OS05,SR98epje} and also for thin polymer films \cite{SJ03}. 
Quantitative applications of classical theories, such as the Rouse or reptation 
models \cite{DoiEdwardsBook}, may be plagued by systematic errors due to the 
neglect of these effective repulsive interactions. They could, for instance, 
be responsible for observed deviations from Rouse behavior \cite{Paul}. Moreover, 
for thin polymer films the repulsive interactions are known to be stronger than in 
the bulk \cite{SJ03} providing, hence, a mechanism to explain the systematic trend 
towards chain swelling observed experimentally for film thicknesses below the bulk 
radius of gyration \cite{JKHBR99nature}.

\acknowledgments
We thank M.~M\"uller (G\"ottingen), M. Rawiso, H.C.~Beno\^it 
(Strasbourg), 
and M.D.~Dadmun (Knoxville) for numerous discussions. 
Computer time by the IDRIS (Orsay) is also gratefully acknowledged.
S.P.O. is indebted to Petroleum Research Grant PRF\# 43923-AC 7 and 
the LEA ``Macromolecules in Finely Divided Media" for financial support. 
J.B. acknowledges financial support by the IUF and from the European 
Community's ``Marie-Curie Actions'' under contract MRTN-CT-2004-504052 
(POLYFILM).

\end{document}